\documentstyle[aps,amssymb,eqsecnum,preprint,tighten,epsf]{revtex}
\begin{document}

\title{A New Way to Make Waves}

\author {	Jeffrey Winicour
       }
\address{Max-Planck-Institut f\" ur
         Gravitationsphysik, Albert-Einstein-Institut, \\
	 14476 Golm, Germany\\
	 and \\
	 Department of Physics and Astronomy \\
         University of Pittsburgh, Pittsburgh, PA 15260, USA 
	 }
\maketitle

\begin{abstract} 
I describe a new algorithm for solving nonlinear wave equations.  In this
approach, evolution takes place on characteristic hypersurfaces. The algorithm
is directly applicable to electromagnetic, Yang-Mills and gravitational fields
and other systems described by second differential order hyperbolic equations.
The basic ideas should also be applicable to hydrodynamics. It is an
especially accurate and efficient way for simulating waves in regions where
the characteristics are well behaved.  A prime application of the algorithm is
to Cauchy-characteristic matching, in which this new approach is matched to a
standard Cauchy evolution to obtain a global solution. In a model problem of a
nonlinear wave, this proves to be more accurate and efficient than any other
present method of assigning Cauchy outer boundary conditions. The approach was
developed to compute the gravitational wave signal produced by collisions of
two black holes. An application to colliding black holes is presented.
\end{abstract} 

\section{Introduction} \label{sec:introduction}

I will describe a new way to ``make waves''. Wave phenomena is treated
mathematically by hyperbolic equations. Einstein's theory of general relativity
is the supreme example. It treats gravity as a distortion in the geometry of
space-time. Gravitational waves are ripples in the geometry which travel
through space changing the shape and size of objects in their path. At this
historic time, Einstein's theory is providing the basis for a new way to
observe the distant universe. A worldwide network of gravitational wave
observatories is under construction. It is designed to detect the gravitational
waves produced throughout the cosmos by collisions between black holes.

The new way to make waves which I will describe was developed to simulate  the
production of gravitational waves by using a computer to solve Einstein's
equations. But the approach applies to any wave phenomena. In order to get
the fundamental ideas across, I present them in terms of very simple
systems. Only near the end will I give in to the urge to tell you about the
application to black holes. However, even in simple examples,
the ideas are easiest to understand in terms of space-time language and
pictures.

The simplest hyperbolic equation is the 1-dimensional advection equation
\begin{equation}
        (\partial_t + c\partial_x) \Phi =0.
\label{eq:advec}
\end{equation}   
It has the general solution $\Phi(t,x)=F(x-ct)$, which describes a wave
traveling with velocity $c$ with initial waveform $F(x)$ at $t=0$. The curves
$x=x_0+ct$ are called the characteristics of the equation. 
They are the paths in space-time (worldlines) along which you must move to
``ride the wave'' in the sense that $\Phi =const$ along a characteristic
worldline. Since $F$ is arbitrary, we can choose $F=0$ for $x<0$ and $F=1$ for
$x>0$ to obtain a wave with a shock front along the characteristic $x=ct$. In
modern parlance, this shock wave carries one bit of information. It illustrates
the general way information is propagated by waves along characteristics
in any hyperbolic system. 

The standard way to solve a hyperbolic equation such as Eq. (\ref{eq:advec})
is by means of the Cauchy problem. One poses initial Cauchy data 
$\Phi(t=0,x)$ in some spatial region $\Omega$ and then evolves the solution in
time. The characteristics causally relate the initial data to a unique
evolution throughout some domain of dependence $D(\Omega)$ in space-time.  As a
result, shock fronts, which represent a sudden signal, can only occur across
characteristics. Hyperbolic equations lead to ordinary differential equations
along the characteristics which govern the propagation of shock
discontinuities. That makes it important for the purpose of numerical
simulation to enforce the propagation along characteristics as extensively as
possible. In complicated cases, where the wave velocity has functional
dependence $c(\Phi,x)$ and an analytic solution is not possible, there still
exists a well established theory of the general properties of hyperbolic
systems based upon characteristics~\cite{john}.  

The method of characteristics~\cite{kreiss} is a computational algorithm for
evolving Cauchy data based upon this theory. As applied to the advection
equation (\ref{eq:advec}), the characteristic velocities are calculated at a
given discretized time $t_n=n\Delta t$ and the algorithm used to determine the
field at time $t_{n+1}$ by requiring that $\Phi$ be constant along the
characteristic curve from a space-time point $P$ at time level $n$ to point
$Q$ at time level $n+1$, i.e.
\begin{equation}
      \Phi_Q=\Phi_P,
\label{eq:1integral}
\end{equation}
where $x_Q =x_P +c\Delta t$. When the wave velocity is not
constant in either space or time, the points $P$ and $Q$ along the
characteristics do not in general lie exactly on spatial grid points, so that
interpolations are necessary. But, because there are no sudden changes {\em
along} characteristics (only {\em across} them), such interpolations give
excellent accuracy by riding the wave like a surfer.

The method of characteristics for the Cauchy problem extends to the
generalization of Eq. (\ref{eq:advec}) to any symmetric hyperbolic system of
first differential order equations in a  multi-dimensional space with many
evolution variables and with source terms for the creation of
waves~\cite{kreiss}. However, there is another classification of hyperbolic
systems, which is more familiar to physicists, based upon second order
differential equations whose principal part has the form
\begin{equation}
     g^{\alpha\beta}\partial_\alpha\partial_\beta \Phi =0,
\label{eq:gwave}
\end{equation} where $\alpha$ is a space-time index, i.e. $x^\alpha=(t,x,y,z)$
for 4-dimensional space-time, and a sum over the values of the repeated indices
is implied. Equation (\ref{eq:gwave}) is classified as hyperbolic if the
symmetric matrix $g^{\alpha\beta}$ has one negative eigenvalue and its
remaining eigenvalues are positive. Displacements $x^\alpha\rightarrow
x^\alpha+dx^\alpha$ along characteristic worldlines
satisfy
\begin{equation}
     g_{\alpha\beta} dx^\alpha dx^\beta  =0,
\label{eq:charact}
\end{equation}
where $g_{\alpha\beta}$, called the metric tensor, is the inverse matrix to
$g^{\alpha\beta}$. The simplest example is the wave equation in one spatial
dimension,
\begin{equation}
       [ \frac {1}{c^2}\partial_t^2 -  \partial_x^2 ]\Phi =0,
       \label{eq:1wave}
\end{equation}
with the displacement along the characteristics satisfying $c^2dt^2-dx^2=0$. By
rewriting Eq. (\ref{eq:1wave}) in the form
\begin{equation}
    (\frac {1}{c}\partial_t -\partial_x) 
         (\frac {1}{c}\partial_t +\partial_x) \Phi =0,
       \label{eq:1wav}
\end{equation}
a comparison with Eq. (\ref{eq:advec}) shows that the general solution is
$\Phi(t,x) =F(x-ct)+G(x+ct)$. There are two characteristics $x=x_0 \pm ct$ 
through each spatial point $x_0$. Information is propagated along these
characteristics which now criss-cross the space-time. The conventional method
of characteristics could be applied to Eq. (\ref{eq:1wav}) by the standard
technique of reducing it to a coupled system of first order differential
equations. But it is simpler to integrate Eq.(\ref{eq:1wav}) over a
parallelogram $\Sigma$ in space time whose sides are characteristics meeting
at the corners $P$, $Q$, $R$ and $S$. This gives the 2-dimensional version of
Eq. (\ref{eq:1integral})
\begin{equation}
     \Phi_Q - \Phi_P + \Phi_R - \Phi_S =0,
\label{eq:2integral}
\end{equation}
as illustrated in Fig. 1 for the case $c=1$ where the characteristics $x=x_0
\pm t$ have $45^o$ slope in space-time. By using Eq. (\ref{eq:2integral}), the
method of characteristics can be implemented as a computational algorithm for
the Cauchy evolution of Eq. (\ref{eq:1wave}) by positioning the point $Q$ at
time level $n+1$, the points $P$ and $S$ at time level $n$ and the point $R$
at time level $n-1$. On a $(t,x)$ computational grid satisfying $\Delta x = c
\Delta t$, the characteristics pass through diagonal grid points. The Cauchy
data, for the wave equation, which consists of the initial values
$\Phi(t=0,x)$ and $\partial_t \Phi(t=0,x)$, is used to initialize an iterative
evolution scheme at time levels $n=0$ and $n=1$. An exceptional feature is
that the one-dimensional wave equation with constant wave velocity can be
evolved {\em without error} by using Eq. (\ref{eq:2integral}) as a finite
difference equation on such a grid. For a variable velocity, the points $P$,
$Q$, $R$ and $S$ cannot be placed exactly on the grid and interpolations are
necessary.

For the wave equation in two or more spatial dimensions, the manner in
which characteristics determine domains of dependence and lead to
propagation equations is qualitatively the same. The major difference is
that an infinite number of characteristics now pass through each point.
For the 3-dimensional wave equation,
\begin{equation}
     [{1 \over c^2}\partial_t^2
      -\partial_x^2-\partial_y^2-\partial_z^2  ]\Phi = 0 \, 
	\label{eq:3d}
\end{equation}
the characteristics which pass through the point $(x_0,y_0,z_0)$ at time
$t_0$ are straight lines which trace out an expanding spherical wavefront
of radius $c(t-t_0)$ at a given time $t$; or, from a space-time point of view,
these characteristics trace out the 3-dimensional characteristic cone 
\begin{equation}
	(x-x_0)^2 + (y-y_0)^2 + (z-z_0)^2 - c^2 (t-t_0)^2 = 0.
	\label{eq:cones}
\end{equation}
In electrodynamics or relativity, the characteristics are light rays and this
is called the light cone. I will often use that language here. The analogue in
hydrodynamics is the Mach cone.

The future (past) light cone consists of the radially outward (inward)
characteristics parameterized by $t>t_0$ ($t<t_0$). There is a 2-parameter
set of characteristics through each point corresponding to the sphere of
angular directions $(\theta,\phi)$ at that point.
This leads to some arbitrariness in formulating an evolution algorithm
for Cauchy data based upon the method of characteristics. There
are an infinite number of characteristics and associated propagation
equations which can be used to evolve Cauchy data from time $t_0$ to
time $t_0+\Delta t$.

For a practical numerical scheme, it is thus necessary
either to average these propagation equations appropriately over the
sphere of characteristic directions at or select out some finite
number of characteristics whose propagation equations comprise a
nonredundant set.  The latter approach has been successfully carried out by
Butler~\cite{butler}.  For the case of plane flow of an inviscid fluid
(a problem in two spatial dimensions), he formulates an
algorithm based upon four ``preferred'' characteristics. In this problem, 
the geometry is further complicated because the characteristics are 
dynamically dependent upon the fluid variables, in contrast to the 
essentially time independent characteristic cones of Eq.~(\ref{eq:cones}).  
In the numerical scheme, the characteristics must themselves be determined 
by some finite difference approximation.

That summarizes the Cauchy problem and its solution by the method of
characteristics.

\section{Characteristic evolution}

I now present a procedure which avoids the arbitrariness and awkwardness of the
method of characteristics in a 3-dimensional Cauchy evolution. It is based upon
a characteristic initial value approach rather than a Cauchy approach. In order
to understand the distinction, it is essential to view space-time as
4-dimensional, with initial data given on a three-dimensional hypersurface. The
Cauchy approach is based upon evolution of initial data given on a spatial
hypersurface in space-time, i.e. points of space at the same instant of time
$t=t_0$, to data at a later time $t=t_0 +\Delta t$. In the characteristic
approach, data is initially posed on an outgoing characteristic cone (light
cone), which defines a hypersurface at constant retarded time $u=u_0$.
Characteristic evolution then proceeds iteratively to characteristic cones
$u_n=u_0+n\Delta u$. The numerical grid is intrinsically based upon these
outgoing characteristic hypersurfaces. The algorithm for evolving from
retarded time $u_n$ to $u_{n+1}$ is based upon characteristics which are
uniquely and intrinsically picked out by the geometry of the retarded time
hypersurfaces.

This characteristic evolution procedure radically differs from Cauchy
evolution. It uses concepts developed in the 1960's for studies of general
relativity~\cite{bondi,null-infinity}. These were prompted by the inability of
the major mathematical tools such as Green's functions and Fourier analysis to
overcome the difficulties posed by nonlinearity of the equations and the
ambiguity of the general coordinate freedom in the theory. Its success later
motivated a more extensive mathematical treatment of the characteristic
initial value problem~\cite{Fried}. These new techniques were especially
designed for investigation of gravitational waves. A computational algorithm
based on this approach  has been the most successful in simulating the
production of gravitational waves from black holes.~\cite{livr,wobb}

The approach has two novel ingredients:
\begin{itemize}
   \item{the use of {\em characteristic hypersurfaces} to formulate a
              characteristic initial value problem (CIVP) and}
   \item{the use of {\em compactification methods} to describe
             on a finite grid the waves propagating to infinity.}
\end{itemize}

{\em Characteristic hypersurfaces} are 3-dimensional sets traced out in
space-time by characteristic worldlines or, equivalently, by their wavefronts.
They provide a natural coordinate system to describe waves~\cite{bondi}. It is
very fruitful to use an initial value scheme which describes  time evolution by
means of the retarded time coordinate defined by characteristic hypersurfaces
as a substitute for the familiar Cauchy scheme based upon constant time
hypersurfaces.  As will be shown, this approach uses a completely different
form of the mathematical equations and the free initial data.

{\em Compactification methods} provide a rigorous description of the radiation
field of a source observed in the asymptotic limit of going to infinity along
a characteristic worldline.\cite{null-infinity}. The key idea is to introduce
a new coordinate which ranges over values from 0 to 1 as the actual distance
from the source ranges from 0 to $\infty$.  The hyperbolic equations are
rewritten in terms of these new coordinates.  Asymptotic behavior at the
``points at infinity'' can then be studied in terms of the new coordinate
which ranges over finite values.  In this way, the concept of the radiation
zone as an {\em asymptotic limit at infinity} is given rigorous meaning. 
Characteristic hypersurfaces are important here since waves travel to infinity
along characteristic worldlines, not along Cauchy hypersurfaces of constant
time. Even for field equations as complicated as those of general relativity,
this procedure provides a finite geometrical description of waves travelling
to infinity. The limit points at infinity form a boundary to the compactified
space-time, which I will refer to as {\em radiative infinity}. At a given
retarded time, this boundary has the topology of a sphere, representing a
sphere of observers at infinity.

It should be emphasized that {\em radiative infinity} differs drastically from
{\em spatial infinity} (the limit of going to infinity holding time constant).
Early considerations of compactifying infinite space for computational
purposes were discarded because they were based upon spatial
infinity~\cite{Orsz}. An attempt to cover infinite space this way by a finite
grid at constant time fails in a  hyperbolic problem because there is
necessarily an infinite physical distance between a grid point at infinity and
its neighbors. This makes it impossible on a spatial grid at fixed time to
resolve radiation with finite wave length which propagates to infinity. In
contrast, for a grid constructed on a characteristic hypersurface,  the grid
points ride the wave without noticing its finite wavelength in the approach to
radiative infinity. 

In terms of characteristic coordinates  $u=ct-x$ and $v=ct+x$,
Eq.~(\ref{eq:1wav}) becomes
\begin{equation}
       \partial_u \partial_v \Phi = 0,
\end{equation}
Here $\Psi = \partial_v \Phi$  satisfies the propagation equation
\begin{equation}
       \partial_u  \Psi  = 0 \label{eq:constraint}
\end{equation}
along the characteristics in the $u$-direction. This is the essence of how the
use of characteristic coordinates simplifies the treatment of waves.

These ideas provide the physical basis for a new computational
algorithm~\cite{Gom}. I will illustrate how it applies to a nonlinear version
of the wave equation in 3 spatial dimensions. However, this evolution algorithm
can be taken over intact to other hyperbolic physical
systems, including electromagnetic fields, the Yang-Mills gauge fields of
elementary particle physics, as well as general relativity, because of the
common mathematical structure of these theories as second differential order
hyperbolic equations. Although it has not been explored how this approach
might be implemented in the case of a first differential order hyperbolic
system, such as hydrodynamics, the general ideas should be applicable.

\section{Characteristic initial value problem for nonlinear waves}

\begin{figure}
\epsfxsize = 14.0cm   \epsfysize = 18.0cm
\centerline{\epsfbox{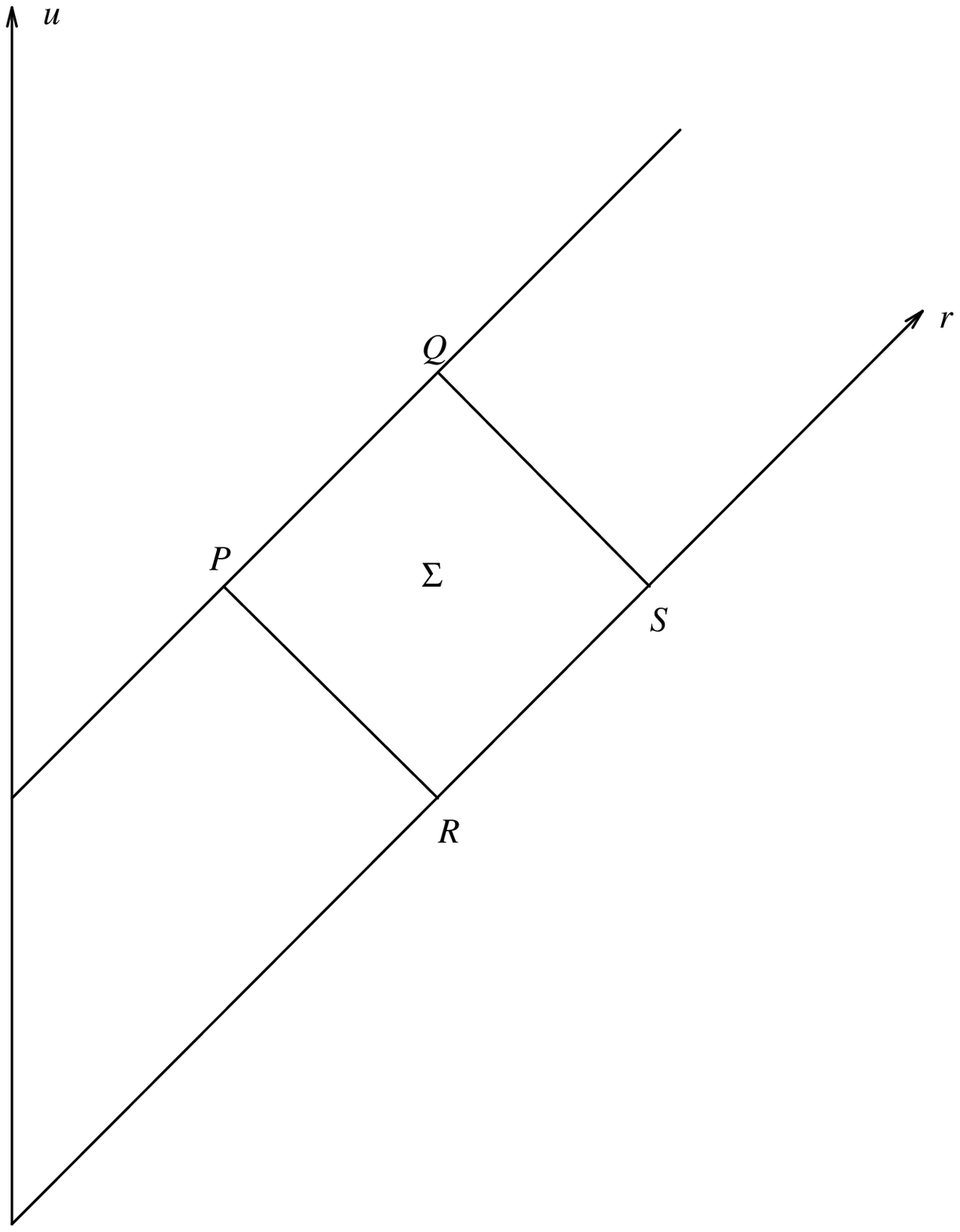}}
\caption{Space-time picture of a parallelogram $\Sigma$ formed by intersecting
characteristics, where the characteristic have $45^o$ slope. The coordinates
are retarded time $u$ and radial distance $r$. In standard $(t,x)$ coordinates,
the $t$-direction is vertical and the spatial $x$-direction horizontal, so that
the pair $P$ and $S$ are at the same time $t$, with $Q$ in their future and $S$
in their past. In $(u,r)$ coordinates, the pair $P$ and $Q$ are at the same
retarded time, as well as the pair $R$ and $S$. } 
\label{fig:parallelogram}
\end{figure}

As a simple illustration of the characteristic initial value problem, consider
the nonlinear scalar wave equation (SWE) in 3-spatial dimensions, which we
write in spherical polar coordinates $(t,r,\theta,\phi)$ as
\begin{eqnarray}
    [{1 \over c^2}\partial_t^2
      -\partial_x^2-\partial_y^2-\partial_z^2  ]\Phi 
      = {1 \over c^2}\partial_t^2 \Phi -{1 \over r}\partial_r^2 (r \Phi)
        + {{L^2 \Phi} \over
       r^2}  = S(\Phi)  \label{eq:PDE}\\
       r = \sqrt(x^2+y^2+z^2),
\end{eqnarray}
where $c$ is the wave velocity, $L^2$ denotes the standard angular
momentum operator
\begin{equation}
	L^2 \Phi = - {{\partial_\theta {(\sin \theta \partial_\theta\Phi)}}
     		  \over {\sin \theta}}
      		  -{{\partial_\phi^2\Phi} \over {\sin^2 \theta}}.
\end{equation}
and $S(\Phi)$ represents a nonlinear source term. Rather than using ordinary
time $t$, characteristic evolution uses the ``retarded time'' coordinate
$u=ct-r$. The outgoing radial characteristics are the curves of constant $u$,
$\theta$ and $\phi$. These are the curves in the $r$-direction, holding
$u=const$, shown in the space-time Fig.~\ref{fig:parallelogram}. In
$(u,r,\theta,\phi)$ coordinates, the SWE (\ref{eq:PDE}) takes the form
\begin{equation}
	2\partial_u\partial_r g =
	 \partial_r^2 g - {{L^2 g} \over r^2}+rS
	 \label{eq:SWE}.
\end{equation}
where $g=r\Phi$.

The striking feature about Eq.(\ref{eq:SWE}) is that it is only first
differential order in {\em retarded time} $u$, unlike the more familiar form of
the SWE (\ref{eq:PDE}) which is of second differential order in {\em time}
$t$.  When data is given on a characteristic initial hypersurface $u=u_0$, we
need only specify the initial value of the field $\Phi$, and then use 
Eq.~(\ref{eq:SWE}) to calculate its retarded time derivative $\partial_u \Phi$
in order to evolve the initial data. This is in contrast to the conventional
Cauchy scheme, where both  $\Phi$ and $\partial_t \Phi$  must be supplied at
initial time $t =t_0$ and Eq.~(\ref{eq:PDE}) is the used to compute the second
time derivative $\partial_t^2 \Phi$ in order to evolve the data.

In a computational implementation of the CIVP, rather than finite differencing
Eq.~(\ref{eq:SWE}) directly, it is advantageous to first convert it into an
integral equation which is subsequently discretized. Using both the outgoing
characteristic coordinate $u$ and the ingoing  characteristic coordinate
$v=ct+r$, the SWE (\ref{eq:SWE}) takes the form
\begin{equation}
     4\partial_u \partial_v g = - {{L^2g} \over  r^2} + r S({g \over r})  
\label{eq:uvSWE}
\end{equation}
where $g = r\Phi$.
In the $u$-$v$ plane formed by fixing the angular coordinates
$(\theta,\phi)$, we construct a parallelogram $\Sigma$ made up of incoming and
outgoing radial characteristics which intersect at vertices $P,Q,R,S$
as depicted in Fig.~\ref{fig:parallelogram}.  By integrating
Eq.~(\ref{eq:uvSWE}) over the area $\Sigma$ bounded by these vertices,
we may establish the identity
\begin{equation}
     g_Q = g_P + g_S - g_R
     +{1 \over  2} \int_\Sigma  du dr  \Bigl [-{{L^2g} \over  r^2} 
        + r S({g \over r}) \Bigr ]    .          \label{eq:integral}
\end{equation}
This simple identity is the 3-dimensional analogue of Eq.
(\ref{eq:2integral}), adapted to include a source term. It
is the starting point for an evolution algorithm which incorporates the 
essential role that characteristics play in the SWE.

In order to study the far field wave behavior, we transform this equation to
the new radial coordinate 
\begin{equation}
	x = r/(1+r), \,\,
        0 \leq x \leq 1.
 \label{eq:coord_trans}
\end{equation}
This serves to map an infinite radial domain into a finite coordinate
region, and assigns infinitely distant radial points to the edge of the
coordinate patch ({\em radiative infinity}) at $x = 1$, where the
radiation signal can be identified.

\subsection{Numerical Algorithm}\label{sec:algorithm}

To develop a discrete evolution algorithm, we work on the
lattice of points
\begin{eqnarray}
	u_n & = & n \Delta u  \\
	x_i & = & i \Delta x  \nonumber \\
       \zeta_{j,k} & = & (j +i k)\Delta \phi   \nonumber  
\end{eqnarray}
where a complex stereographic coordinate $\zeta$ is used to cover the
the sphere in two patches (North and South) in order to avoid the
polar singularities of the spherical coordinates $(\theta,\phi)$.
We denote the field at these sites by
\begin{equation}
	g^n_{ijk} =  g(u_n, x_i, \zeta_{j,k}).
\end{equation}
(We will generally suppress the angular indices $j$ and $k$.)

\begin{figure}
\epsfxsize = 14.0cm   \epsfysize = 18.0cm
\centerline{\epsfbox{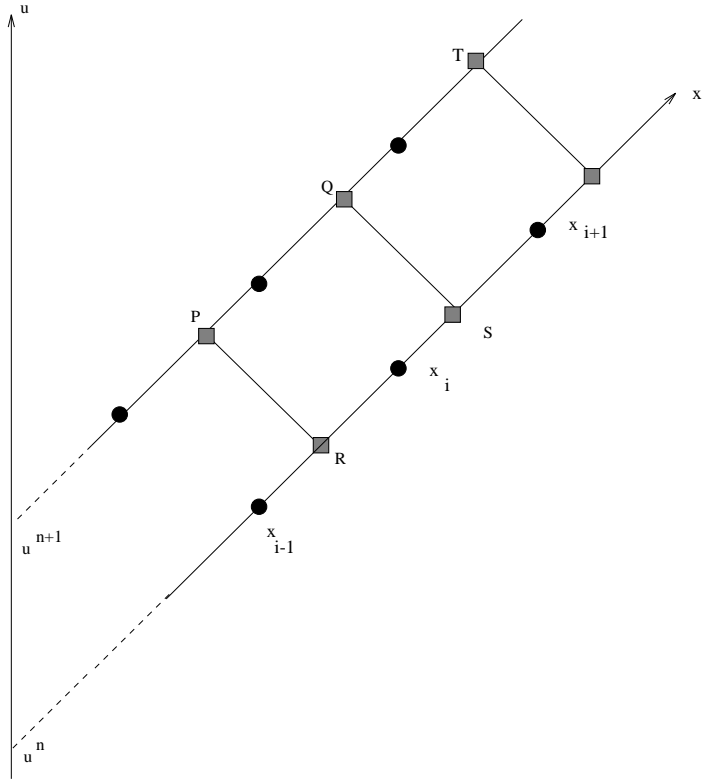}}
\caption{Line segments drawn at forty-five degrees represent radial
characteristics. Their intersection defines the fundamental null parallelogram
PQRS shown superimposed upon the computational cell, which consists of the
points marked by circles and their nearest neighbors in the angular directions
(not shown). Here $u$ is the retarded time coordinate and  $x$ is the
compactified radial coordinate. The marching algorithm determines $\Phi_Q$ in
terms of the previously determined values $\Phi_P$, $\Phi_R$ and $\Phi_S$. The
process is then iterated to determine $\Phi_T$ on a march to radiative
infinity. }
\label{fig:cell}
\end{figure}

With respect to the $(u,x)$ coordinate grid, it is not possible to place the
corners P, Q, R and S at  grid points since the slope of the characteristics in
the compactified $x$ coordinate depends upon location. As a consequence, the
field $g$ at these points must be interpolated from neighboring grid points.
The essential feature of the placement of the parallelogram on the grid is that
the sides formed by the ingoing characteristics intersect adjacent
$u$-hypersurfaces at equal but opposite $x$-displacements from the neighboring
grid points, as illustrated in Fig.~\ref{fig:cell}. The field values at the
vertices of the parallelogram are obtained by quadratic interpolation.
Cancellations between the interpolation errors at the four vertices yields the
accuracy
\begin{equation}
	g_Q-g_P-g_S+g_R = G_Q-G_P-G_S+G_R +O((\Delta x)^3\Delta u), 
\end{equation}
where $G$ represents the exact analytic solution. 

The integral in Eq.~(\ref{eq:integral}) can be evaluated by treating 
the integrand as a constant over the parallelogram, with value at the center.   The radial coordinate of the point at
the center is $r_c = (r_P + r_S)/2$.  To compute the
nonlinear term, the value of $g$ at $r_c$ is taken as the average $g_c
= (g_P +g_S)/2$, with $g_P$ and $g_S$ evaluated from second-order
linear interpolations over adjacent points on the grid.  The angular
derivatives in Eq.~(\ref{eq:integral}) are replaced with standard
second-order-accurate finite difference approximations. $L^2g$ is
calculated on the grid points, and the same interpolation procedure is
used to obtain the value of $L^2g_c$. The integral term is then approximated by
\begin{eqnarray}
     \int_\Sigma[ -L^2g + r^3 S(g/r)] du \, dr /r^2
        = [ - L^2g_c + r_c^3 S(g_c/r_c)]\int_\Sigma du \, dr/r^2 \nonumber\\
        = 2 \log \bigl ({{r_Q r_R} \over {r_P r_S}} \bigr )
         \bigl [ - L^2g_c + r_c^3 S({g_c \over r_c}) \bigr ] \ , 
\end{eqnarray}
where the integrand is accurate to second order in $\Delta x$
and $\Delta \phi$.   
The resulting finite difference equation
\begin{eqnarray}
\lefteqn{ g^{n+1}_{i}(x_Q+x_P-x_{i-2}-x_{i-1}) = }      \nonumber \\
        & & 2g^{n+1}_{i-1}(x_Q+x_P-x_{i-2}-x_{i})       
            -g^{n+1}_{i-2}(x_Q+x_P-x_{i-1}-x_{i})       \nonumber \\
        & & + \{ g^{n}_{i+1}(x_S+x_R-x_{i-1}-x_{i}) 
            -2g^{n}_{i}(x_S+x_R-x_{i-1}-x_{i+1})        \nonumber \\     
        & & +g^{n}_{i-1}(x_S+x_R-x_{i}-x_{i+1}) \} 
           { {(x_S-x_R)} \over {(x_Q-x_P)} }            \nonumber \\
        & & +\int_\Sigma { {dr du} \over r^2 } 
            \Bigl [ - L^2(g_c) + r_c^3 S({g_c \over r_c}) \Bigr ] 
            { (\Delta x)^2 \over {(x_Q-x_P)} } .        
\label{eq:algorithm}
\end{eqnarray}
relates values of $g^{n+1}_i$ with values at neighboring grid points which are
either earlier in retarded time ($g^n_{i-1}, g^n_i, g^n_{i+1}$), or else
contemporary but located at smaller radius ($g^{n+1}_{i-2},g^{n+1}_{i-1}$).
Consequently, it is possible to move through the grid, computing $g^{n+1}_i$
explicitly by an orderly march. This is achieved by starting at the origin at
time $u_{n+1}$. Field values of $g=r\Phi$ vanish there. Step outward to the
next radial point, using Eq.~(\ref{eq:algorithm}) for all angular sites on the
grid, and iterate this march out to radiative infinity thus updating the
characteristic cone at $u_{n+1}$ and  completing one retarded time step. This
march is then iterated in retarded time.

The algorithm steps $g$ radially outward one cell with a local error of fourth
order in grid size. This leads to second order global accuracy which is
confirmed by convergence tests using known analytic solutions~\cite{Gom}. A
complete specification of the algorithm would require a description of how the
startup procedure at the origin is handled and how stereographic coordinates
are used to compute angular derivatives in a smooth way. Details of the use of
North and South stereographic grids are given in Ref.~\cite{competh}. Physical
behavior of nonlinear waves is treated in Ref.~\cite{Gom}. Construction of an
exact nonspherical solution for an $S=\Phi ^3$ self-interaction allows 
calibration of the algorithm in the nonlinear case where physical singularities
form. The predicted second order accuracy is confirmed right up to the
formation of the singularity. Other choices of nonlinear potential allow
simulation of solitons.

The Courant-Friedrichs-Lewy (CFL) condition that the numerical domain of
dependence contain the physical domain of dependence (determined by the
characteristics) is a necessary condition for convergence of a finite
difference algorithm. For a grid point at $(u,r,\theta)$, there are three
critical grid points, at $(u-\Delta u,r+\Delta r,\theta)$ and $(u-\Delta
u,r-\Delta r,\theta \pm \Delta \theta)$, which must lie inside its past
characteristic cone. These gives rise to the inequalities $\Delta u < 2\Delta
r$ and $\Delta u < -\Delta r+(\Delta r^2 +r^2 \Delta \theta^2)^{1/2}$. At large
$r$, the second inequality becomes  $\Delta u < r\Delta \theta$ and the Courant
limit on the time step is essentially the same as for a Cauchy evolution
algorithm. However, near the vertex of the cone, the second inequality gives a
stricter condition
\begin{equation} 
      \Delta u < K\Delta r \Delta \theta^2, 
\label{eq:cfl} 
\end{equation}
where the value of $K$ depends upon the start up procedure at the vertex. For
the scalar wave equation, these stability limits were confirmed by numerical
experiments~\cite{Gom} and it was found that $K\approx 4$.

Local von Neumann stability analysis leads to no constraints on the algorithm.
This may seem surprising because no analogue of a CFL condition on
the time step arises. It can be understood in the following vein. The local
structure of the code is implicit, since it involves 3 points at the upper
time level.  Implicit algorithms do not necessarily lead to a CFL condition.
However, the algorithm is explicit in the way that the evolution starts up
as an outward radial march from the origin. It is this startup procedure that
introduces a CFL condition.

Operating within the CFL limit, the algorithm gives a stable, globally second
order accurate evolution on a compactified grid~\cite{Gom}. (In some nonlinear
applications, artificial dissipation is necessary for stability~\cite{lehner}.)
Radiative infinity behaves as a perfectly transmitting boundary so that no
radiation is reflected back into the system. Numerical evolution  satisfies a
conservation law relating the loss of energy to the radiation flux at
infinity. 

That summarizes the characteristic initial value problem and its implementation
as a new computational approach to simulate waves.  

\section{Cauchy-characteristic matching}

Characteristic evolution has many advantages over Cauchy evolution. Its one
disadvantage is caused by the existence of either (i) caustics where
neighboring characteristics focus or (ii), a milder version of this,
cross-over points where two distinct characteristics collide. The vertex of
the characteristic cone is a highly symmetric example of a point caustic where
a complete sphere of characteristics focus. I have already discussed how a
point focus gives rise to a strong limitation imposed by the CFL condition. 

Cauchy-characteristic matching (CCM) is a way to avoid such limitations by
combining the strong points of characteristic and Cauchy evolution in
formulating a global evolution.~\cite{livr,Bis2,prl96}  Here I illustrate the
application of CCM to the nonlinear wave equation (\ref{eq:PDE}). This problem
requires boundary conditions at infinity which ensure that the total energy
and the energy loss due to radiation are both finite. In a 3-dimensional
problem, these are the conditions responsible for the proper $1/r$ asymptotic
decay of the radiation fields.  However, for practical purposes, in the
computational treatment of such a system by the Cauchy problem, an outer
boundary is artificially established at some large but finite distance.  Some
condition is then imposed upon this boundary in an attempt to approximate the
proper asymptotic behavior at infinity. Such an artificial boundary condition
(ABC) typically causes partial reflection of the outgoing wave back into the
system~\cite{Orsz,Lind,Hig86,Ren}, which contaminates the accuracy of the
evolution and the radiated signal.  Furthermore, nonlinear wave equations
often display backscattering so that it may not be correct to try to entirely
eliminate incoming radiation from the numerical solution. The errors
introduced by ABC's are of an analytic origin, essentially independent of the
computational discretization. In general, a systematic reduction of the error
can only be achieved by simultaneously refining the grid and moving the
computational boundary to a larger radius, which is computationally very
expensive for three-dimensional simulations. CCM provides a global solution
which does not introduce error at the analytic level.

For linear wave problems, a variety of ABC's have been proposed. For recent
reviews, see Ref's. \cite{Ren,giv,tsy,ryab}. During the last two decades,
local ABC's in differential form have been extensively employed by several
authors \cite{Lind,Hig86,Eng77,Bay80,Tre86,Bla88,Jia90} with varying success.
Some local ABC's have been derived for the linear wave equation by considering
the asymptotic behavior of outgoing solutions \cite{Bay80}; this approach may
be regarded as a generalization of the Sommerfeld outgoing radiation
condition. Although such ABC's are relatively simple to implement and have a
low computational cost, their final accuracy is often limited because their
simplifying assumptions are rarely met in practice \cite{giv,tsy}. Systematic
improvement of the accuracy of local ABC's can only be achieved by moving the
computational boundary to a larger radius.

The disadvantages of local ABC's have led to implementation of nonlocal ABC's
based on integral representations of the infinite domain problem
\cite{giv,tsy,Tin86}. Even for problems where the Green's function is known and
easily computed, such approaches were initially dismissed as impractical
\cite{Eng77}; however, the rapid increase in computer power has made it
possible to implement nonlocal ABC's for the linear wave equation even in 3
space dimensions \cite{deM}. For a linear problem, this can yield numerical
solutions which converge to the exact infinite domain problem as the grid is
refined, keeping the artificial boundary at a fixed distance. However, due to
nonlocality, the computational cost per time step usually grows at a higher
power of grid size ($O(N^4)$ per time step in a 3-dimensional problem with
$O(N^3)$ spatial grid points) than in a local approach \cite{giv,tsy,deM},
which is demanding even for today's supercomputers. Further, the applicability
of current nonlocal ABC's is restricted to problems where nonlinearity may be
neglected near the grid boundary \cite{tsy}.

To my knowledge, only a few works have been devoted to the development of ABC's
for strongly nonlinear problems~\cite{giv,Tho87,Hag88}. In practice, nonlinear
problems are often treated by linearizing the governing equations in the far
field , using either local or nonlocal linear ABC's \cite{tsy,ryab}. Besides
introducing an approximation at the analytical level, this procedure requires
that the artificial boundary be placed sufficiently far from the strong-field
region, which sharply increases the computational cost in multidimensional
simulations.  There seems to be no currently available ABC which is able to
produce numerical solutions which converge (as the discretization is refined)
to the infinite domain exact solution of a strongly nonlinear 3-dimensional
wave problem, keeping the artificial boundary at a fixed location.

For such nonlinear problems, CCM produces an accurate solution out to
radiative infinity with effort $O(N^3)$ per time-step. CCM increases the total
computational cost only by a factor $\sim 2$ with respect to a pure Cauchy
algorithm with a local ABC. The use of numerical methods based upon matching a
characteristic initial-value formulation and a Cauchy formulation can
effectively remove the above difficulties associated with a finite
computational boundary. There is no need to truncate space-time at a finite
distance from the sources, since compactification of the radial coordinate
makes it possible to cover the whole space-time with a finite grid.  In this
way, the true radiation zone signal may be computed. Although the
characteristic formulation has stability limitations in interior region where
the characteristic hypersurfaces can develop caustics, it proves to be both
accurate and computationally efficient in the treatment of the exterior,
caustic-free region.

CCM is a new approach to global numerical evolution which is free of error at
the analytic level.  The characteristic algorithm provides the {\em outer}
boundary condition for the interior Cauchy evolution, while the Cauchy
algorithm supplies the {\em inner} boundary condition for the characteristic
evolution. Since CCM consists of discretizing an exact analytic treatment of
the radiation from source to radiative infinity, it generates numerical
solutions which converge to the exact analytic solution of the radiating system
even in the presence of strong nonlinearity.  Thus, any desired accuracy can be
achieved by refining the grid, without moving the matching boundary. In
practice, the method performs extremely well even at moderate resolutions.

\subsection{CCM for nonlinear waves}

As an illustration of the computational implementation of CCM , consider again
the nonlinear 3-dimensional wave equation (\ref{eq:PDE}). For simplicity, set
the velocity $c = 1$. In the standard computational implementation of the
Cauchy problem for (\ref{eq:PDE}), initial data $\Phi(t_0,x,y,z)$ and
$\partial_t\Phi(t_0,x,y,z)$ are assigned and evolved in a bounded spatial
region, with some ABC imposed at the computational boundary. In a
characteristic initial-value formulation the SWE (\ref{eq:PDE}) is reexpressed
in the form of Eq. (\ref{eq:SWE}) in terms of $g=r\Phi$, using standard
spherical coordinates and a retarded time coordinate $u = t - r$. The initial
data $g(u_0,r,\theta,\phi)$, on an initial outgoing characteristic cone $u
=u_0$ is then evolved globally out to radiative infinity.

In CCM, (\ref{eq:PDE}) is solved in an interior region $r \le R_{m}$ using a
Cauchy algorithm, while a characteristic algorithm solves the retarded
coordinate version (\ref{eq:SWE}) for $r \ge R_{m}$. The matching procedures
ensure that, in the continuum limit, $\Phi$ and its gradient are continuous
across the interface $r = R_{m}$. This is a requirement for any consistent
matching algorithm, since a discontinuity in the field or its gradient could
act as a spurious boundary source, contaminating both the interior and exterior
evolutions.

For technical simplicity, I illustrate the details of the method here for
spherically symmetric waves but it has been successfully implemented in fully
nonlinear 3-dimensional problems without symmetry.  In a 3-dimensional problem
without symmetry, the characteristic evolution is carried out on an exterior
spherical grid, while the Cauchy evolution uses a Cartesian grid covering the
interior spherical region.  Although a spherical grid could also be used in the
interior, a Cartesian grid avoids the necessity of cumbersome numerical
procedures to handle the singularity of spherical coordinates at the origin. A
Cartesian discretization in the interior and a spherical discretization in the
exterior are the coordinates natural to the geometries of the two regions.
However, this makes the treatment of the interface somewhat involved; in
particular, guaranteeing the stability of the matching algorithm requires
careful attention to the details of the inter-grid matching. Nevertheless,
there is a reasonably broad range of discretization parameters for which CCM is
stable~\cite{ccm}.

\begin{figure}
\epsfxsize = 15.0cm   \epsfysize = 15.0cm
\centerline{\epsfbox{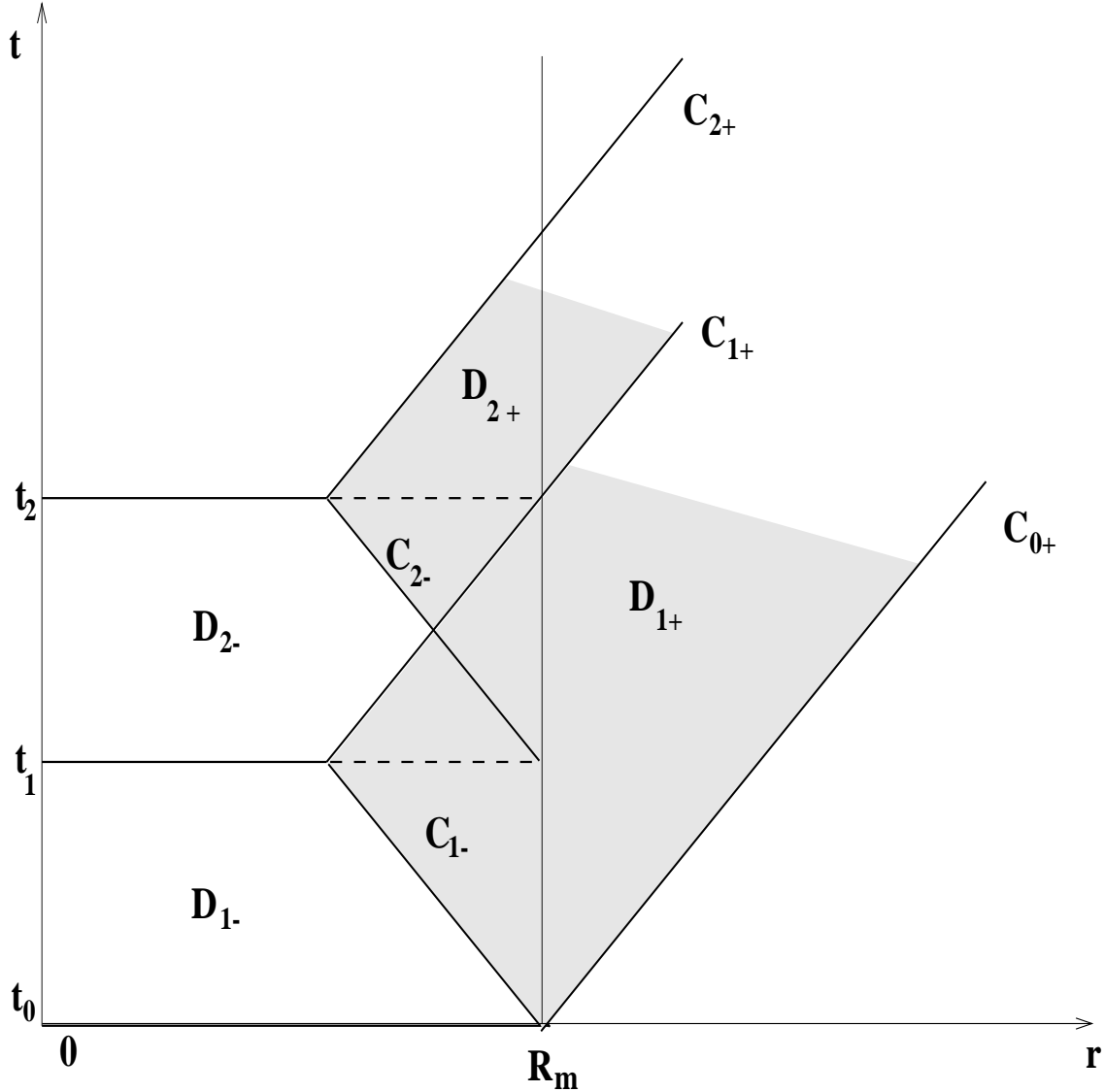}}
\caption{Initial Cauchy data are evolved from $t_0$ to time $t_1$
throughout the region $D_{1^-}$. Characteristic data induced on
$C_{1^-}$, combined with the initial characteristic data on $C_{0^+}$
are used to evolve the region $D_{1^+}$.  This produces Cauchy data at
time $t_1$ in the region $r\le R_{m}$.  Similarly, Cauchy evolution is used
in the region $D_{2^-}$, bounded on the right by  $C_{2^-}$. The
characteristic data induced on $C_{2^-}$, together with those on
$C_{1^+}$, are sufficient to evolve through the region $D_{2^+}$. The process
can be iterated to carry out the entire future evolution of the
system.}
\label{fig:1d-locked}
\end{figure}

With the substitution $G=r\phi$ and the use of spherical coordinates, the
spherically symmetric version of the SWE (\ref{eq:PDE}) reduces to the
1-dimensional wave equation
\begin{equation}
    \partial_{tt}G = \partial_{rr}G + rS. \label{eq:1dswe}
\end{equation}
The initial Cauchy data is $G(t_0,r)$ and $\partial_t G(t_0,r)$ in the region
$0\le r\le R_{m}$.  Together with the regularity condition $G(t,0)=0$, these
data determine a unique solution in the domain of dependence $D_{1^-}$
indicated in Fig.~\ref{fig:1d-locked}.  The outer boundary of the domain of
dependence is the ingoing radial characteristic $C_{1^-}$ described by
$r=R_{m}-t+t_0$. The solution cannot be constructed throughout the complete
interior region $r \leq R_{m}$ without additional information, which can be
furnished by giving  the value of $G$ on the outgoing characteristic $C_{0^+}$
described by $r=R_{m}+t-t_0$ (see Fig.~\ref{fig:1d-locked}). In terms of the
coordinates $u=t-r$ and $r$, $C_{0^+}$ is described by $u=t_{0}-R_{m}$ In these
coordinates, expressing $g(u,r)=G(u+r,r)$, the spherically symmetric version of
the wave equation (\ref{eq:SWE}) is
\begin{equation}
	2\partial_{ur}g = \partial_{rr}g +rS.
        \label{eq:1dSWE}
\end{equation}
A unique solution of (\ref{eq:1dSWE}) is determined by characteristic initial
data consisting of the value of $g$ on the initial outgoing characteristic
$C_{0^+}$ and on the ingoing characteristic $C_{1^-}$. These data determine the
solution uniquely throughout the future of $C_{1^-}$ and $C_{0^+}$, i.e. the
region $D_{1^+}$ in Fig.~\ref{fig:1d-locked}.

The matching scheme proceeds as shown in Fig.~\ref{fig:1d-locked}.
First, initial Cauchy data are evolved from $t_0$ to $t_1$ throughout the
region $D_{1^-}$, which is in its domain of dependence. Next, this induces
characteristic data on $C_{1^-}$ which combined with the
initial characteristic data on $C_{0^+}$ allows a characteristic
evolution throughout the region $D_{1^+}$, bounded in the future by the
characteristic $C_{1^+}$.  The solution determined from this initial
stage induces Cauchy data at time $t_1$ in the region $r\le R_{m}$,
inside the matching boundary. This process can then be iterated
to carry out the entire future evolution of the system.

\begin{figure}
\epsfxsize = 15.0cm   \epsfysize = 15.0cm
\centerline{\epsfbox{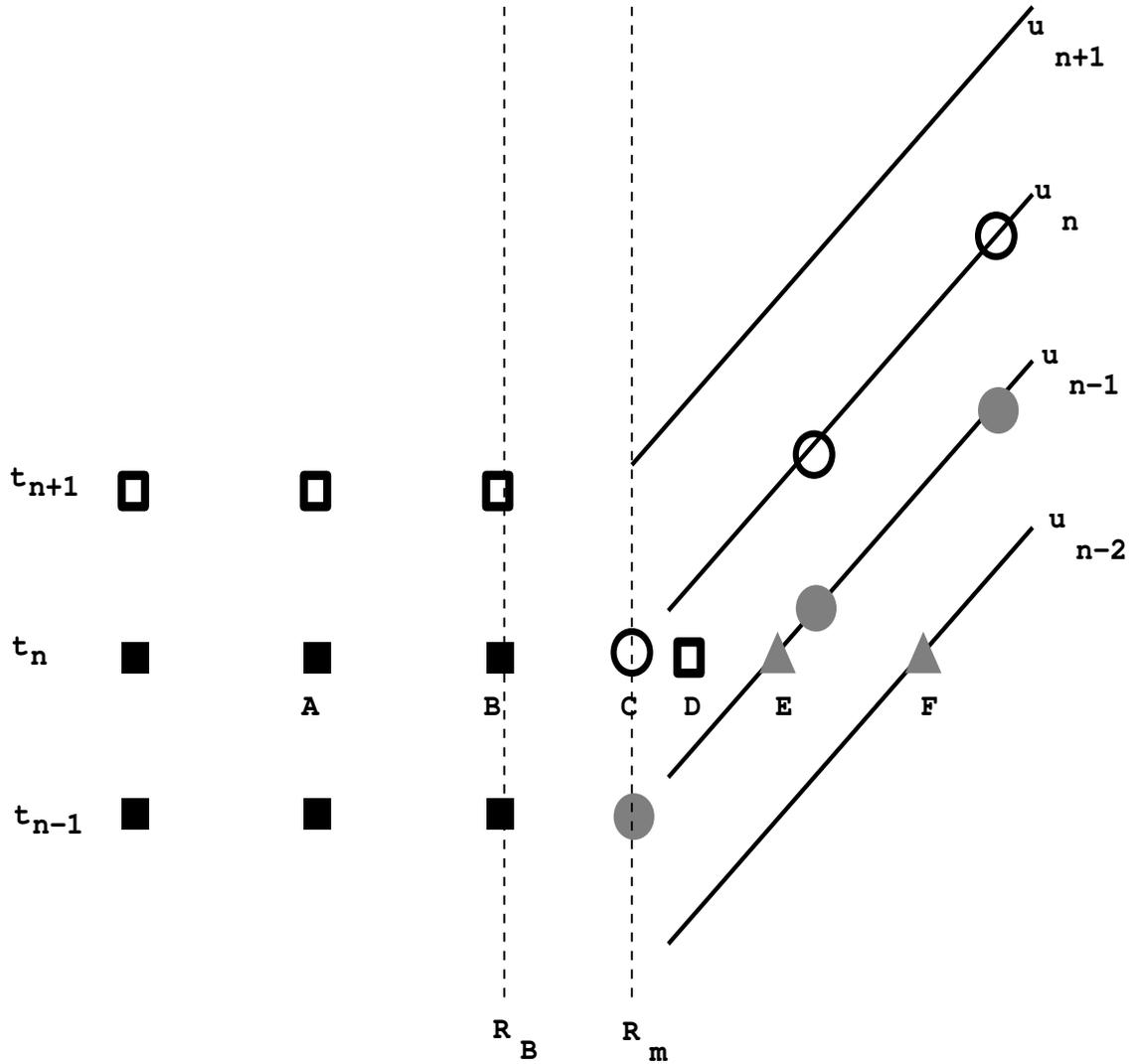}}
\caption{Cauchy grid points are indicated by squares and characteristic
grid points by circles. The triangles indicate points $E$ and $F$ where
the time level $t_n$ intersects the retarded time levels $u_{n-1}$ and
$u_{n-2}$. Initial data are given at the shaded points. Evolution
proceeds iteratively by determining field values at the unshaded
points.  The matching scheme provides boundary values at $C\ (r =
R_{m})$ and $D\ (r = R_{B} + \Delta r)$ for the characteristic and Cauchy
grids, respectively.}
\label{fig:1d-match}
\end{figure}

CCM is a discretized version of this scheme in which the criss-cross
pattern of characteristics inside the radius $R_{m}$ is at the scale of a grid
spacing. The discretized evolution algorithm consists of the following steps
(see Fig.~\ref{fig:1d-match}):

{\em Step 1. Cauchy evolution.}
The interior integration scheme is implemented on a uniform spatial grid
$r_i = i\Delta r$ ($0\le i \le M$) with outer radius
$R_B = M\Delta r$. We discretize (\ref{eq:1dswe}) using the
standard second-order finite difference scheme
\begin{equation}
      {G_i^{n+1}-2G_i^n +G_i^{n-1} \over (\Delta t)^2} = {G_{i+1}^n
     -2G_i^n +G_{i-1}^n \over (\Delta r)^2} + r_i S_i^n,
      \label{eq:1cauchy}
\end{equation}
where $G_{i}^{n} = G(t_{n},r_{i})$, $S_{i}^{n} = S(t_{n},r_{i})$, and $t_{n} =
t_{0} + n\Delta t$. The interior evolution is initialized by evaluating
$G_{i}^{0}$ and $G_{i}^{1}$ $(0\leq i\leq M)$ to second order accuracy from the
Cauchy initial data. In the $n$th time step, (\ref{eq:1cauchy}) is used to
compute $G_{i}^{n+1},\ 1\le i \le M$ in terms of field values $G_{i}^{n-1}$ and
$G_{i}^{n}$. The regularity of $\Phi$ at $r = 0$ implies that $G_{0}^{n} = 0$
for all $n$. The boundary values $G_{M+1}^{n}$ which are required by
(\ref{eq:1cauchy}) are supplied by the matching procedure (step 3).

{\em Step 2. Characteristic evolution.}
The characteristic algorithm is implemented on a uniform grid based
on the dimensionless compactified radial coordinate
\begin{equation}
  \eta=1-{1\over 1+r/R_{m}} \,\, \, ,
        \frac{1}{2}\le \eta \le 1 \label{eq:w}
\end{equation}
so that points at radiative infinity (corresponding to $\eta=1$) are included
in the grid. In order to include one of the technical problems in matching
an interior Cartesian grid to an exterior spherical grid, let there be a small
gap between the outer radius $R_B$ of the Cauchy grid and the matching radius
$R_m$ (which is also the inner radius of the characteristic grid). In
3-dimensional Cartesian-spherical matching the outermost Cauchy grid points
and the innermost characteristic grid points are necessarily distinct. This is
represented here by a gap $R_{m} - R_{B} = \kappa\Delta r$, where $\kappa \ge
0$ is an arbitrary parameter. The characteristic grid consists of the
uniformly spaced points $\eta_{\alpha} = \frac{1}{2}+\alpha\Delta\eta$ ($0
\leq \alpha \leq N_{\eta}$), where $\Delta\eta =(2N_\eta)^{-1}$.  The retarded
time levels $u = u_{n}$ for the characteristic evolution are chosen so that $u
= u_{n}$ intersects the time level $t = t_{n}$ of the Cauchy evolution at the
matching radius; therefore, $u_n=t_{n}-R_{m}$ and $\Delta u=\Delta t$.  We
denote by $g_{\alpha}^{n}$ the value of $g$ at $\eta = \eta_{\alpha}$, $u =
u_{n}$. The initial characteristic data consist of $g^{0}_{\alpha}$, $0 \le
\alpha \le N_{\eta}$.

In the $n$th iteration of the evolution, we compute the field values at the
grid points with $u = u_{n}$ using the values of $g^{n-1}_{\alpha}$, which are
known either from initialization or from the previous iteration. As already
described, this is done by the characteristic marching algorithm based on the
integral identity (\ref{eq:integral}).

At the inner boundary of the characteristic grid ($\alpha = 1$), the previous
scheme must be slightly modified, since $g^{n}_{\alpha-2}$ is not defined.
For this initial step, $PQRS$ is chosen so that $\eta_{P} = \eta_{0}$,
$\eta_{Q} = \eta_{1}$, and $g_{R}$, $g_{S}$ are approximated by quadratic
interpolation in terms of $g^{n-1}_{0}$, $g^{n-1}_{1}$, $g^{n-1}_{2}$, which
have already been computed. Besides these field values, the final evaluation
of $g^{n}_{1}$ still requires the value of $g^{n}_{0}$, which is supplied
by the matching procedure (step 3).

{\em Step 3. Matching.}
Numerous schemes are possible in the case of spherically symmetric
matching. Here I describe one which is stable for a
wide range of gap sizes, $0\le\kappa\le2$ and works in
the more complicated situation with an interior Cartesian grid and an
exterior spherical grid.

The required boundary values $G_{M+1}^{n}$ and $g^{n}_{0}$ are computed
by radial interpolations at constant $t$, using the field values at
points $A$, $B$, $E$, and $F$ in Fig.~\ref{fig:1d-match}. The first two of
these field values are already known at the $n$th step, while the last two
can be obtained by cubic radial interpolations along the previously evolved
characteristics $u = u_{n-1}$ and $u = u_{n-2}$, respectively. At the
initial step, point $F$ lies on the characteristic $u = u_{-1} = -\Delta t -
R_{m}$, which is not evolved by the algorithm; this field
value is supplied along with the initial data. Once the Cauchy and
characteristic boundary values are computed, a new iteration may be
performed starting from Step 1 above.

Since all the interpolations employed in the matching step have fourth order
error, the matching algorithm has the same second order global accuracy
exhibited by the separate Cauchy and characteristic algorithms, as confirmed by
numerical tests.

In summary, CCM has been implemented and tested for nonlinear waves in
3-dimensional space. No special assumption is made about the waves crossing
the computational interface $r = R_{m}$ and nonlinear effects in the exterior
characteristic domain are automatically taken into account. In numerical
experiments CCM  converged to the exact solution (with the matching boundary
fixed at an arbitrary position) in highly nonlinear problems. For
comparison, nonlocal ABC's yielded convergent results only in linear
problems.  In terms of both computational cost and accuracy, CCM is a very
effective way to solve the 3-dimensional wave equations, with or without a
nonlinear term. It is possible to achieve convergence with an ABC by refining
the grid {\em and simultaneously} enlarging the radius of the outer boundary. 
However, this is very expensive computationally, especially for small target
error in the determination of the radiated signal in a 3-dimensional
problems~\cite{cce}. Because CCM is convergent under grid-refinement alone,
for small target error its performance is significantly better than any
available alternative. In strongly nonlinear problems. CCM appears to be the
only available method which is able to produce numerical solutions which
converge to the exact solution with a computational interface located at an
arbitrary fixed position.

\section{Application to black holes}

By definition, a black hole traces out a world tube in space-time which
is the boundary of what is visible to an outside observer. As a
result, this worldtube is necessarily a characteristic hypersurface traced
out in space-time by light rays (the characteristics of general relativity).

A mechanical analogue of a black hole can in principle be constructed by
letting a reservoir of water empty through a funnel into a lower reservoir. By
arranging the flow velocity in the funnel to exceed the velocity of sound in
the upper reservoir, an acoustic version of a black hole results~\cite{unruh}.
Sound waves can travel from the upper reservoir to the lower but not in the
reverse direction.

The simplest example of a black hole arises in the spherically symmetric
collapse of a star whose energy has been depleted to the extent that internal
pressure cannot withstand gravity. A point sized black hole first forms at the
center of the collapsing star, which then expands into a sphere growing with
the velocity of light, thus tracing out a light cone in space-time. Normally a
light cone keeps expanding forever. What makes a black hole light cone unique
is that gravity halts this expansion and eventually the black hole just hovers
in equilibrium at a fixed size. Each point on the spherical black hole still
moves along a light ray, but the sphere of light rays is in a delicate
gravitational balance between growing and shrinking.

This spherically symmetric black hole was discovered analytically by
Schwarzschild as a simple solution to Einstein's equations. Unfortunately,
Schwarzschild's black hole does not emit gravitational waves (just as a
spherically symmetric charge distribution does not emit electromagnetic waves).

The inspiral and merger of a binary system of black holes is a powerful source
of gravitational waves, which lie in the frequency band detectable by the new
gravity wave observatories. The lack of symmetry necessitates a computational
treatment to determine the waveform of the radiated signal. The binary black
holes trace out a characteristic hypersurface in space-time so that their
simulation by characteristic evolution is a natural approach. Using a
characteristic evolution algorithm similar to that which I have described here,
except now applied to Einstein's equation, we have found fascinating results.
The binary black holes, rather than initially forming as a point caustic in the
Schwarzschild case, form as a cross-over surface where light rays collide. This
cross-over surface is itself bounded by a ring of caustics which in a sense mark
the merger of the two individual black holes into a single black
hole.~\cite{ndata,asym}

\begin{figure}
\epsfxsize = 10.0cm   \epsfysize = 8.0cm
\centerline{\epsfbox{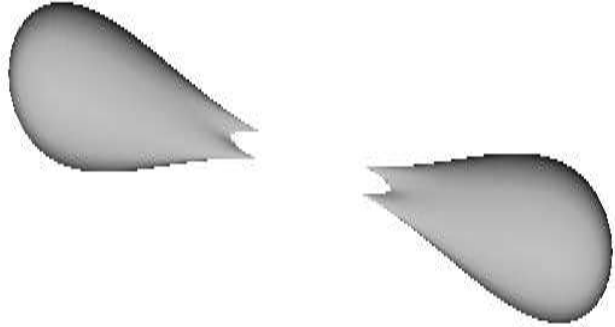}}
\caption{Prelude to a generic collision of black holes. As they
approach, the initially spherical black holes are tidally distorted.}
\vspace*{10pt}
\label{fig:cantor}
\end{figure}

\begin{figure}
\epsfxsize = 10.0cm   \epsfysize = 8.0cm
\centerline{\epsfbox{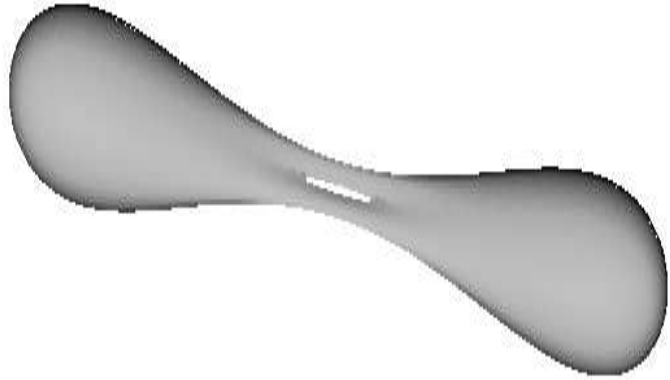}}
\caption{Temporarily toroidal black hole produced by the merger of two black
holes. The hole in the torus soon closes up to form a peanut shaped black hole,
which then expands into a spherical shape.}
\vspace*{10pt}
\label{fig:kytor}
\end{figure}

The individual black holes form as spheres but, as illustrated in Fig.
\ref{fig:cantor}, as the holes approach, their mutual gravitational tidal
distortion produces sharp pincers just prior to merger. At merger, the pincers
join to form a single temporarily toroidal black hole, as illustrated in Fig.
\ref{fig:kytor}. The inner hole of the torus subsequently closes up to produce
first a single peanut shaped black hole and finally a spherical black hole. 
Details of this merger can be viewed at the web site
http://artemis.phyast.pitt.edu/animations. We are now using characteristic
evolution to compute the gravitational wave signal emitted in such black hole
collisions in the anticipation that they will be detected by the new gravity
wave observatories.

\section{Acknowledgements}

I thank the Pittsburgh Supercomputing Center and NPACI for making computing
time available for this research and  the National Science Foundation for
research support under grant NSF PHY 9510895.

\end{document}